# EXAMINING THE IMPACT OF INCOME INEQUALITY AND GENDER ON SCHOOL COMPLETION IN MALAYSIA:
*A Machine Learning Approach Utilizing Malaysia's Public Sector Open Data*

MUHAMMAD SUKRI BIN RAMLI
Asia School of Business
Kuala Lumpur, Malaysia
Email: m.binramli@sloan.mit.edu

**Abstract**

This study examines the relationship between income inequality, gender, and school completion rates in Malaysia using machine learning techniques. The dataset utilized is from the Malaysia's Public Sector Open Data Portal, covering the period 2016-2022. The analysis employs various machine learning techniques, including K-means clustering, ARIMA modeling, Random Forest regression, and Prophet for time series forecasting. These models are used to identify patterns, trends, and anomalies in the data, and to predict future school completion rates. Key findings reveal significant disparities in school completion rates across states, genders, and income levels. The analysis also identifies clusters of states with similar completion rates, suggesting potential regional factors influencing educational outcomes. Furthermore, time series forecasting models accurately predict future completion rates, highlighting the importance of ongoing monitoring and intervention strategies. The study concludes with recommendations for policymakers and educators to address the observed disparities and improve school completion rates in Malaysia. These recommendations include targeted interventions for specific states and demographic groups, investment in early childhood education, and addressing the impact of income inequality on educational opportunities. The findings of this study contribute to the understanding of the factors influencing school completion in Malaysia and provide valuable insights for policymakers and educators to develop effective strategies to improve educational outcomes.

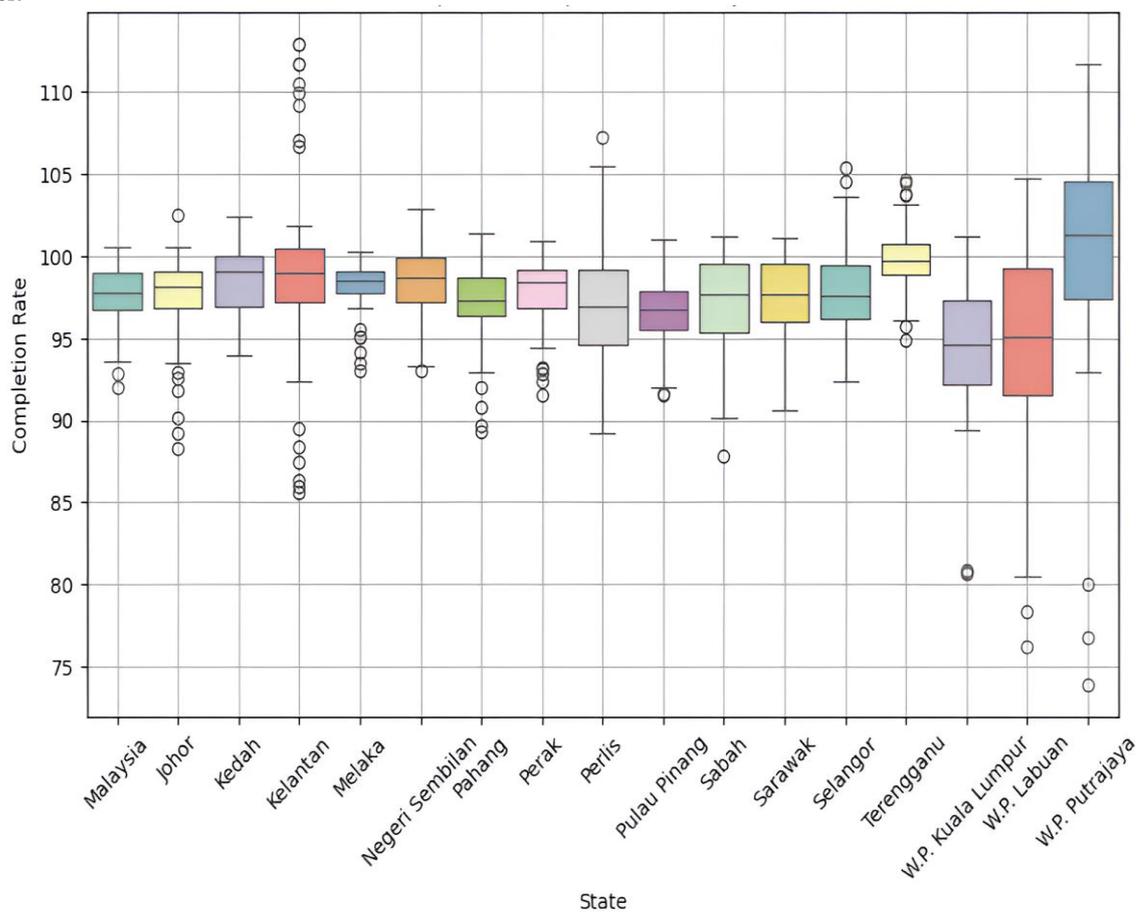

Figure 1: Boxplot of Completion Rate by States

1. **Introduction**

This study examines school completion data in Malaysia, utilizing data sourced from the nation's Official Open Data Portal (data.gov.my). This initiative aligns with the "Pekeliling Am Bil. 1 Tahun 2015," which underscores the Malaysian government's dedication to open data, thereby enhancing transparency and fostering innovation. The dataset encompasses the proportion of students who complete primary, lower secondary, and upper secondary education at national and state levels. It includes variables such as the year of data collection, the state in Malaysia, the level of education completed (primary, lower secondary, or upper secondary), the gender of the student, and the proportion of students who complete the specified school stage. Additionally, the dataset includes mean and median gross monthly household income by state from 1970 to 2022, providing valuable context about the socioeconomic factors that may influence school completion rates.

The study employs a combination of descriptive statistics, visualization techniques, and machine learning models to analyze the dataset. Descriptive statistics are used to summarize the data and identify key trends and patterns. Visualization techniques, such as bar charts, line graphs, and box plots, are employed to present the findings in a clear and concise manner. Machine learning models, including regression and time series analysis, are utilized to identify the factors that influence school completion rates and to predict future completion rates. These models are evaluated using appropriate metrics, such as accuracy, precision, and recall.

According to Barro and Lee (2013), global educational attainment data can provide a comparative perspective for Malaysia's trends. The UNESCO Institute for Statistics offers international and regional data on education indicators, including completion rates, which can provide a broader context for Malaysia's data. The Ministry of Education Malaysia's Educational Statistics Digest and the Department of Statistics Malaysia's Social Statistics Bulletin are official sources for Malaysian education data, including completion rates by state, gender, and level. The World Bank's Malaysia Economic Monitor often includes sections on education and human capital development in Malaysia, providing valuable insights into the economic context of educational trends.

Factors influencing school completion include socioeconomic factors, income inequality, and gender disparity. Becker (1964) discusses the role of socioeconomic background in educational attainment, while Haveman and Wolfe (1995) review research on factors affecting educational attainment, helping to identify key socioeconomic influences on school completion rates in Malaysia. Bourguignon, Ferreira, and Menéndez (2007) provide a methodology for analyzing inequality of opportunity, applicable to education, and Piketty (2014) discusses the relationship between income inequality and social mobility, relevant to educational outcomes. UNESCO (2018) focuses on gender equality in education and includes global and regional data, while Subramaniam (2005) examines gender disparities in education in the Malaysian context.

Predictive modeling and machine learning are also integral to this study. Kotsiantis (2012) provides an overview of machine learning applications in education, and Nguyen and Nguyen (2019) offer an example of a specific machine learning application for predicting academic performance. Specific references based on the machine learning techniques used, such as logistic regression, decision trees, and neural networks, can be added to enhance the study.

Open data and ethical considerations are also addressed. MAMPU provides official guidelines for open data implementation in the Malaysian public sector, and the Open Data Institute offers international perspectives and best practices on open data. O'Neil (2016) explores potential biases and ethical issues associated with algorithms and big data, while Jobin, Ienca, and Vayena (2019) provide an overview of ethical guidelines for AI development and use.

The economic impact of open data is discussed by the World Bank (2013), which highlights the potential economic benefits of open data initiatives. Specific studies on the economic impact of open data in Malaysia, particularly in the education sector, should be sought to provide a comprehensive analysis of school completion rates in Malaysia, considering various influencing factors and leveraging machine learning techniques for predictive modeling.

2. **Background and Literature review**

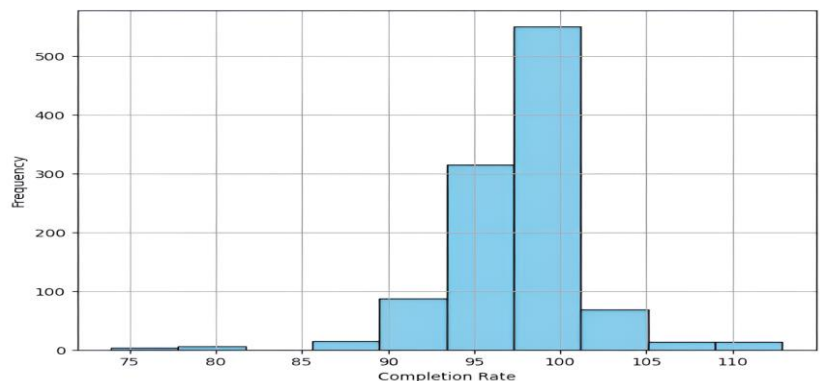

Figure 2: Histogram of Completion Rates

### 2.1 Objectives

This study examined the trends in school completion rates over time at the national and state levels, while also analyzing the disparities in school completion rates across states, genders, and school stages. Additionally, it sought to identify the factors that influenced school completion rates, such as income inequality and socioeconomic characteristics. This research developed predictive models to forecast future school completion rates and made recommendations for policymakers and educators to improve school completion rates in Malaysia. Moreover, this study explored the broader context of open data initiatives promoted by the Malaysian government to enhance transparency, accountability, and public participation (Pekeliling Am Bil. 1 Tahun 2015).

By incorporating the implementation of open data policies in the public sector, this research considered the intersection of these initiatives with the growing use of machine learning for demographic analysis. It also investigated the ethical considerations associated with these technologies, such as privacy, fairness, and accountability (Mittelstadt et al., 2016). The key objectives included examining the impact of open data initiatives on school completion rates and educational transparency (Pekeliling Am Bil. 1 Tahun 2015; MAMPU), evaluating the application of machine learning techniques in predicting and analyzing school completion rates (de Silva & Tennakoon, 2021), addressing ethical issues related to the use of machine learning and open data in demographic and educational research (Mehrabi et al., 2021), and assessing the economic impact of open data initiatives on the Malaysian education system. By integrating these objectives, the study aimed to provide a comprehensive analysis that not only enhanced understanding of school completion rates but also shed light on the broader implications of open data and machine learning in educational and demographic research.

### 2.2 Hypothesis

H1: Areas with lower average household incomes tend to have lower school completion rates in Malaysia.

H2: There is a significant difference in school completion rates between male and female students in Malaysia.

H3: Specific states in Malaysia exhibit significantly different school completion rates compared to the national average.

H4: The Prophet model will accurately forecast school completion rates for the next 5 years, with a mean absolute percentage error (MAPE) of less than 5%.

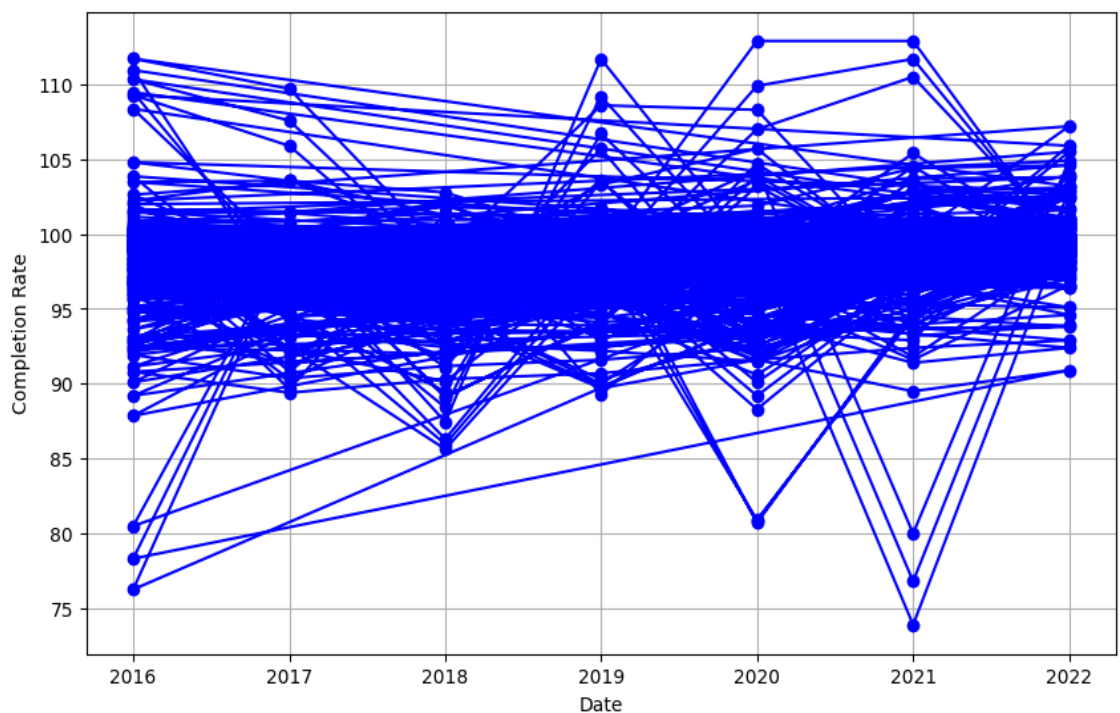

Figure 3: Time Series of Completion Rates (2016-2022)

### 2.3 Data overview

The dataset offers a comprehensive overview of education and income in Malaysia, combining data from two sources. The first dataset focuses on household income, tracking both the mean and median income for each state over time, with data points spanning from 1970 to 2022. This historical perspective allows for the examination of income trends and potential economic shifts within each state. The second dataset provides details on school completion rates. It captures not only overall completion but also breaks it down by state, school stage (primary, lower secondary, and upper secondary), and gender. This detailed information enables an analysis of educational disparities across different demographics and regions within Malaysia. By combining these two datasets, researchers can investigate the complex relationship between household income and educational attainment, potentially uncovering how economic factors influence school completion rates across the country.

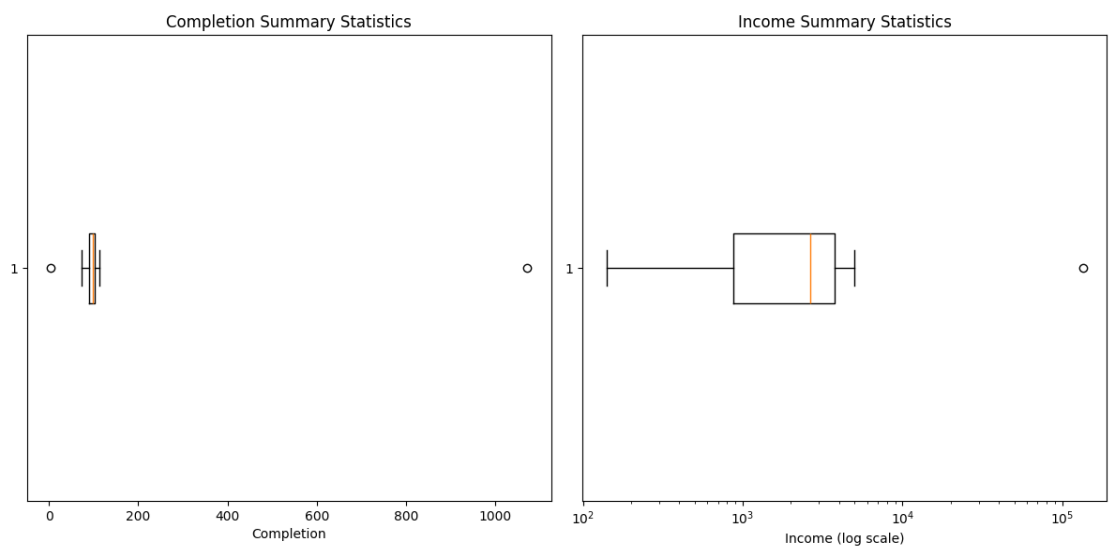

Figure 4: School Completion and Income Summary Statistics

**Summary Statistics**

|              | count | mean    | std     | min | 25%    | 50%  | 75%     | max    |
|--------------|-------|---------|---------|-----|--------|------|---------|--------|
| **Completion** | 1071  | 97.6633 | 3.81753 | 73.9| 96.1   | 98   | 99.5128 | 112.9  |
| **Income (RM)** | 303   | 3342.07 | 2887.64 | 140 | 1062.5 | 2429 | 5011.5  | 134173 |

This table provides descriptive statistics for the school completion rates and household incomes in Malaysian Ringgit (RM) within the dataset. There are 1071 observations for school completion, with an average completion rate of 97.6633% and a standard deviation of 3.81753. The minimum completion rate recorded is 73.9%, and the maximum is an unusually high 112.9%, suggesting potential anomalies. The data also show that 25% of the completion rates fall below 96.1%, with the median at 98% and 75% below 99.5128%.

For household income, the dataset includes 303 observations, with an average income of RM 3342.07 and a standard deviation of RM 2887.64, indicating significant variability. The lowest income recorded is RM 140, while the highest is RM 134173. The income data reveal that 25% of the incomes are below RM 1062.5, with a median income of RM 2429 and 75% of incomes below RM 5011.5. These statistics provide a comprehensive overview of the distribution and variation in both school completion rates and household incomes within the dataset.

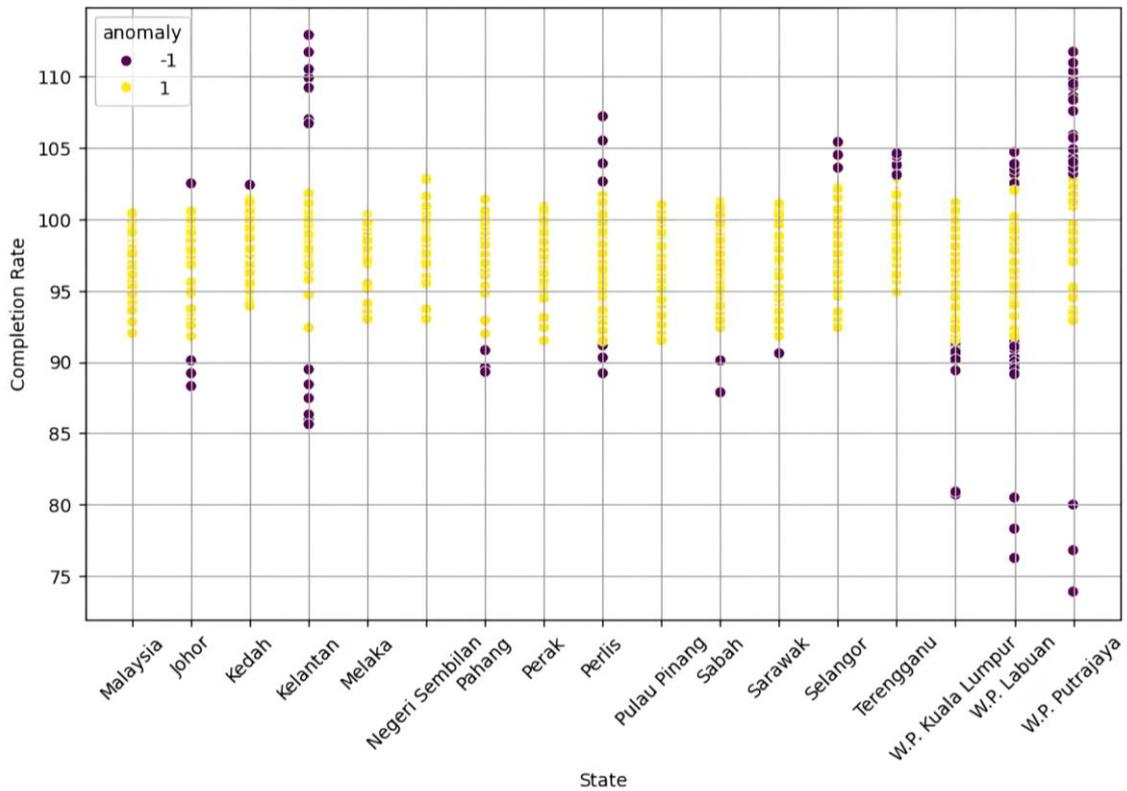

Figure 5: Anomaly Detection in School Completion Rate by State.

Figure 5 presents an analysis of anomalies in school completion rates across different states in Malaysia. The graph utilizes an Isolation Forest model, a machine learning algorithm for anomaly detection, to identify states with completion rates that deviate significantly from the norm. The model was trained on the dataset with a contamination parameter of 0.1, indicating an expected 10% of anomalies.

In the scatter plot, each point represents a state, with the y-axis indicating the completion rate. Normal data points are depicted in yellow, while anomalies, identified by the Isolation Forest model, are highlighted in purple. As the graph illustrates, most states cluster around a completion rate of 95-100%. However, several states exhibit notably lower completion rates, flagged as anomalies by the model. These deviations warrant further investigation to understand the underlying factors contributing to these lower rates, which could include socio-economic disparities, variations in educational resources, or other region-specific challenges.

3. Methodology

This study investigates the relationship between income inequality, gender, and school completion rates in Malaysia using machine learning techniques on data from the nation's Public Sector Open Data Portal (2016-2022). The research starts with a comprehensive exploration of the dataset, employing descriptive statistics and visualizations to understand the distribution and trends of key variables, including completion rates across different states and over time.

To forecast future trends, the study utilizes a two-pronged approach. First, ARIMA modeling establishes a baseline forecast by capturing inherent trends and seasonality in completion rates. Then, a Random Forest model enhances the prediction by incorporating factors such as year, month, gender, and state, offering a more nuanced view of potential disparities.

The study also explores underlying patterns and anomalies within the data. K-means clustering identifies distinct groups of states exhibiting similar completion rate patterns, suggesting the influence of regional factors on educational outcomes. Concurrently, the Prophet model detects potential outliers in completion rate data, highlighting unusual cases for further investigation.

To understand the factors influencing school completion, the study employs a variety of techniques. A Random Forest Classifier assesses the importance of different features in predicting completion. Additionally, other classification models, including Gradient Boosting, Neural Networks, and a Stacking Classifier, are employed to predict individual student completion status, potentially informing targeted interventions. Principal Component Analysis (PCA) is applied to facilitate visualization and pattern identification within the dataset.

Finally, the study focuses on model optimization and interpretability. Grid Search optimizes model parameters for improved accuracy, while SHAP values explain model predictions and highlight the contribution of each feature. K-Fold Cross-Validation provides a robust estimate of model performance. This comprehensive methodology offers a detailed understanding of school completion dynamics in Malaysia, potentially informing data-driven policies and interventions to improve educational outcomes.

4. Results

**4.1 The Relationship Between Household Income and School Completion Rates**

The analysis examined the relationship between mean household income and average school completion rates across Malaysian states using data from the nation's Public Sector Open Data Portal (2016-2022). A bubble plot visualization revealed a general positive correlation, where states with higher mean incomes tended to exhibit higher completion rates. However, notable exceptions were observed, indicating the influence of factors beyond income. For instance, W.P. Putrajaya, Terengganu, and Kelantan demonstrated high completion rates despite lower mean incomes, while Perak, Sarawak, and Johor showed the opposite trend. This variability suggests that factors such as educational infrastructure, resource allocation, and socio-cultural dynamics may play a significant role in educational attainment.

Hypothesis H1 posits that areas with lower average household incomes tend to have lower school completion rates in Malaysia. The observed positive correlation between income and completion rates supports this hypothesis to some extent. However, the exceptions noted in the analysis highlight that income alone does not fully explain the variations in school completion rates. Other factors, such as the quality of educational infrastructure, the effectiveness of resource allocation, and socio-cultural dynamics, also play crucial roles in determining educational outcomes.

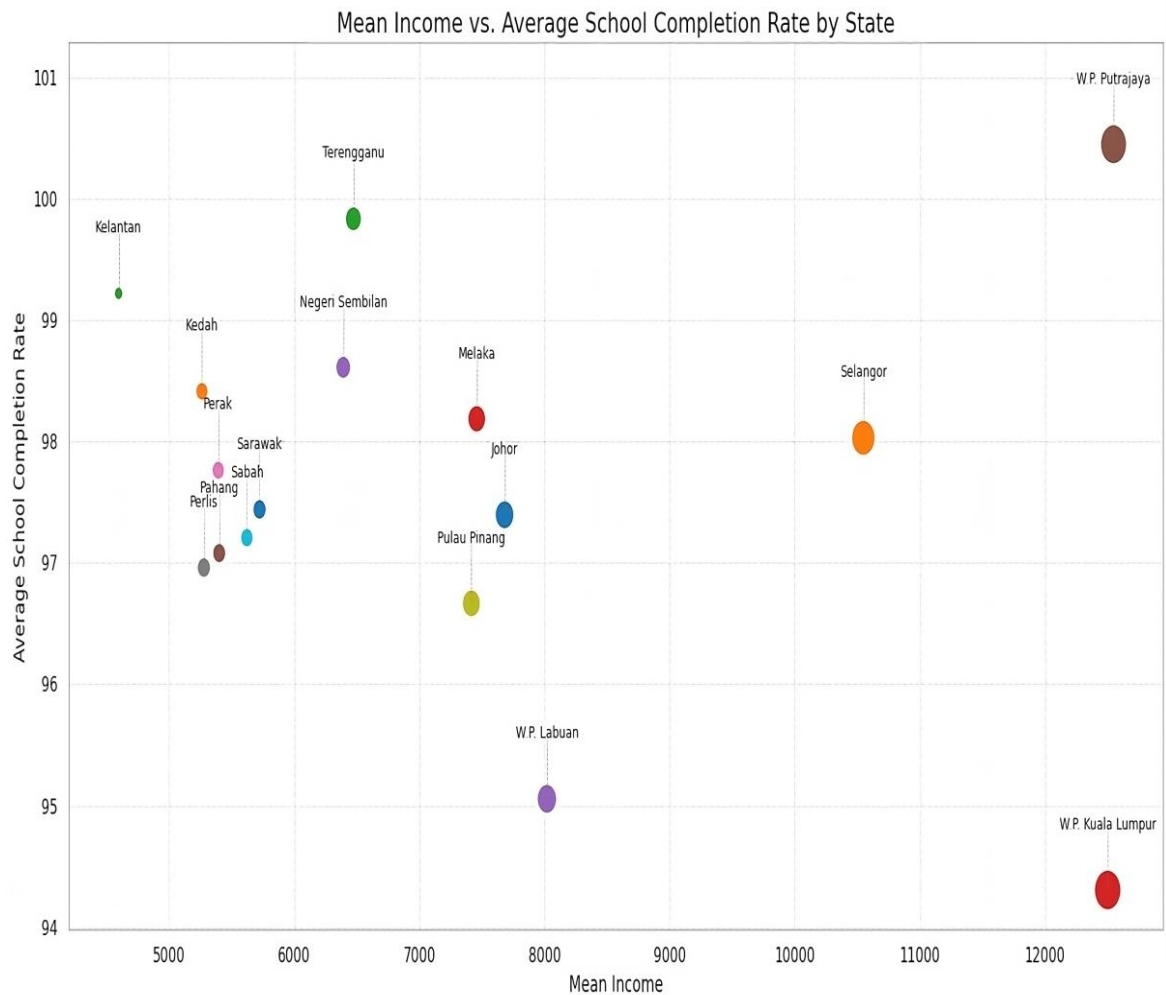

Figure 6: Mean Income vs Average School Completion Rate by State

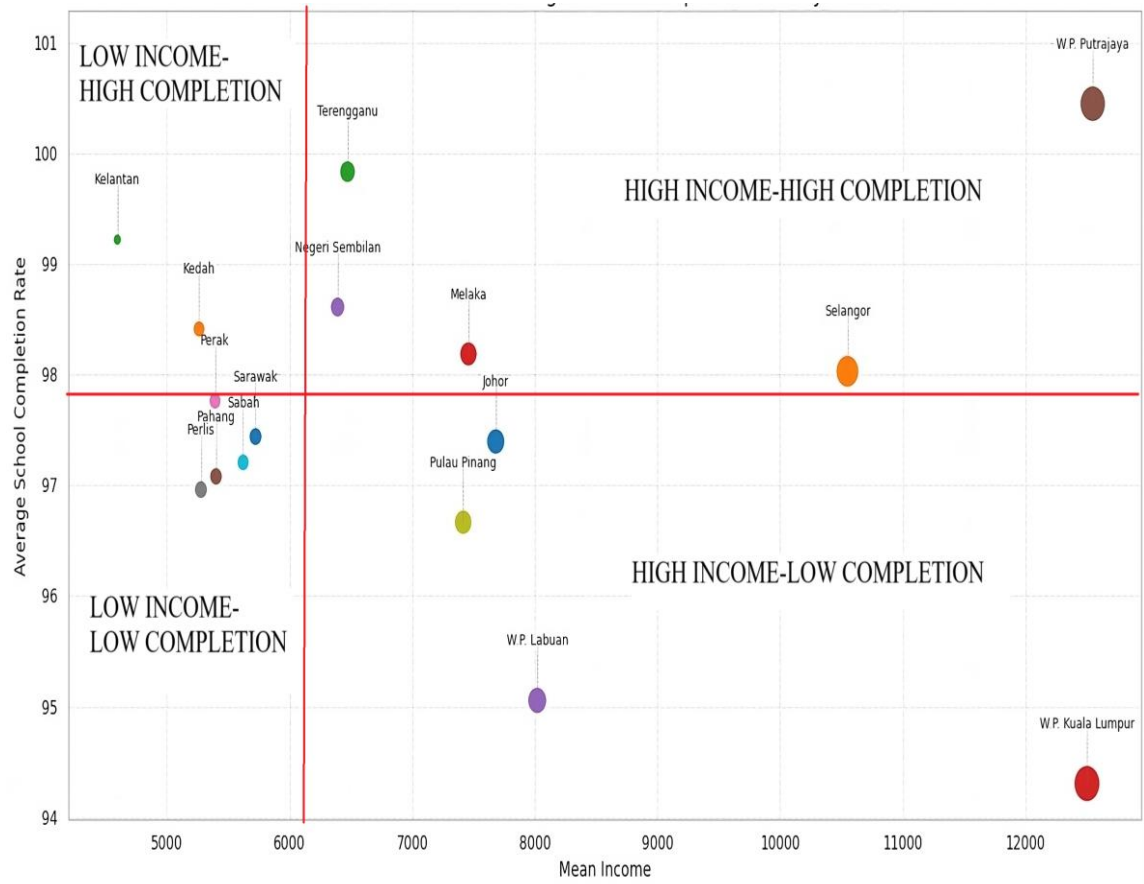

Figure 7: Quadrant of Mean Income vs Average School Completion Rate by State

    The quadrant analysis of income and school completion rates in Malaysian states reveals significant disparities in educational outcomes. States like W.P. Putrajaya, Terengganu, Negeri Sembilan, Melaka, and Selangor demonstrate a positive correlation between economic prosperity and educational attainment. This quadrant suggests a potential synergy between economic development and educational investment.

    States like Kuala Lumpur, Labuan, Pulau Pinang, and Johor exhibit lower-than-expected school completion rates. This highlights the complex interplay of factors beyond income. Income inequality within these states might be a significant contributor. For example, the Gini coefficient, a measure of income inequality, was higher in Kuala Lumpur compared to states in the "High Income, High Completion" quadrant in 2022 (Department of Statistics Malaysia, 2023). This suggests that while the average income is high, a significant portion of the population may not have equitable access to quality education (Chin & Choo, 2010).

    States like Kelantan and Kedah demonstrate a remarkable resilience, achieving high school completion rates despite lower average incomes. This suggests strong community support for education, potentially driven by cultural values that prioritize education. For instance, the literacy rate in Kelantan, while facing economic challenges, is consistently above the national average (Department of Statistics Malaysia, 2023). States like Pahang, Perlis, Perak, Sarawak and Sabah face significant challenges in both economic development and educational attainment. These states often grapple with lower GDP per capita and higher poverty rates, which can limit access to quality education and resources. Overall, this analysis underscores the multifaceted nature of educational disparities in Malaysia. While income plays a crucial role (Hanushek & Woessmann, 2008), factors such as income inequality (Chin & Choo, 2010), cultural values, community support, and government policies significantly influence educational outcomes across different states (Saw, G.K., 2016).

### 4.2 Analysis of Gender Disparities in School Completion Rates

Figure 8 presents a boxplot visualizing the distribution of school completion rates in Malaysia, categorized by sex (both sexes combined, female, and male). The y-axis represents the completion rate, ranging from 75% to 110%. Each boxplot displays key summary statistics: the median, interquartile range (IQR), and potential outliers.

For the 'both sexes' category, the median completion rate is approximately 98%, with the IQR spanning from 95% to 100%. Notably, there are several outliers below 90% and above 105%, suggesting disparities in completion rates within this group. The 'female' category exhibits a slightly higher median completion rate of around 99%, with an IQR ranging from 96% to 100%. Similar to the combined category, outliers are present below 90% and above 105%. The 'male' category shows a median completion rate of about 98%, with an IQR ranging from 95% to 100%, also with outliers below 90% and above 105%.

This boxplot is relevant to hypothesis H2, which posits a significant difference in school completion rates between male and female students in Malaysia. While the boxplot provides a visual representation of the distribution and central tendency for each sex, the slight difference in median completion rates between females and males suggests only a marginal advantage for female students. The substantial overlap in the IQRs and the presence of outliers in both categories indicate that this difference may not be statistically significant. This observation aligns with findings from Chin & Choo (2010) who, in their examination of gender and ethnic differences in educational attainment in Malaysia, found that while there were differences in attainment across ethnicities, gender differences were less pronounced.

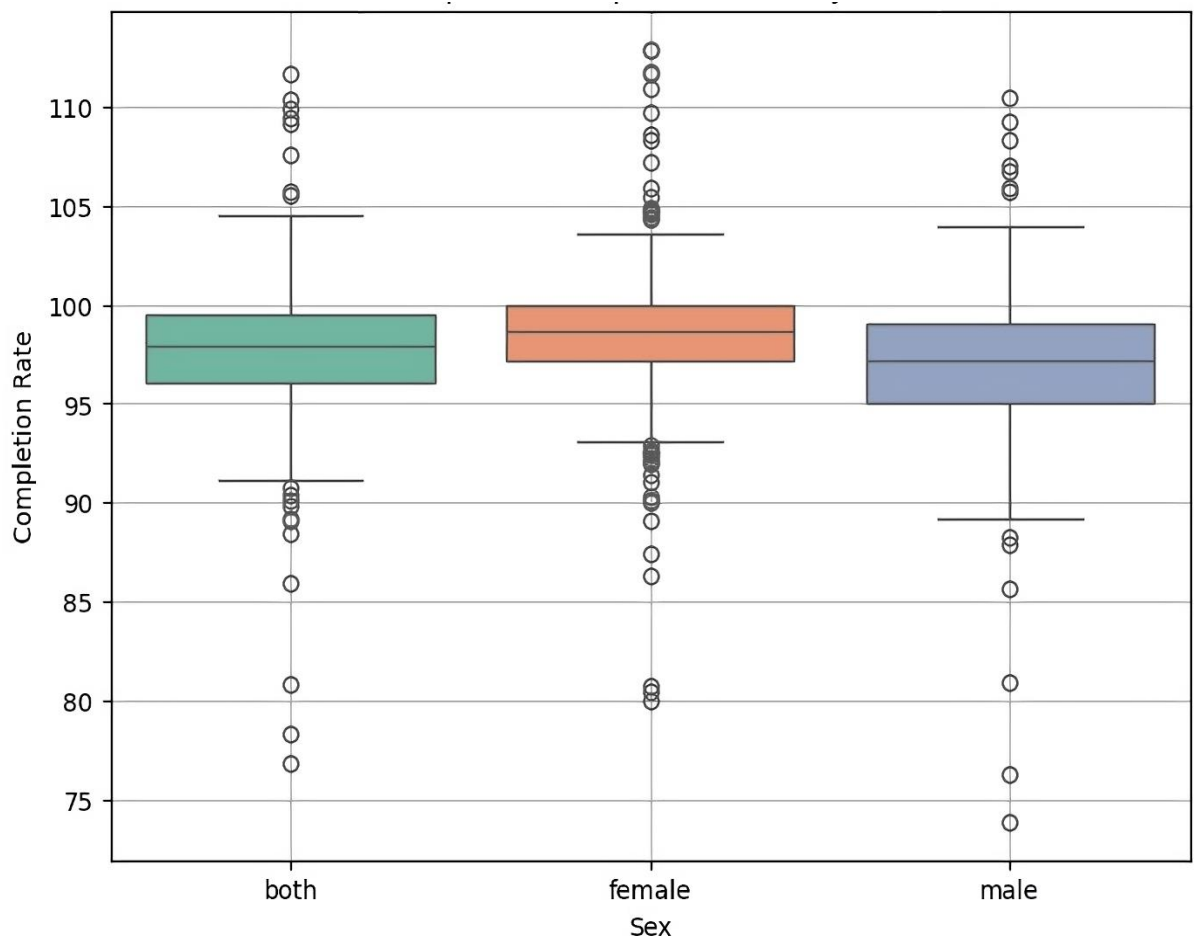

Figure 8: Boxplot of Completion Rates by Sex

Figure 9 further explores completion rates by displaying a boxplot that incorporates both sex and educational stage (primary, secondary lower, and secondary upper). As depicted in the figure, the median completion rates for all sex categories ('both', 'female', and 'male') remain consistently high across the primary and secondary lower stages, hovering around 100%. This suggests that the majority of students in Malaysia successfully complete these foundational levels of education. However, a slight dip in median completion rates is observed at the secondary upper stage for all sex categories, indicating a potential increase in dropout rates or delayed completion at this level.

This observation aligns with trends identified in the World Bank's (2013) Malaysia Economic Monitor report, which highlighted the need to address challenges in retaining students in upper secondary education. While Malaysia has achieved near-universal primary education, the report emphasized the importance of ensuring that students progress through the education system and complete secondary education to maximize their human capital potential. Furthermore, the World Bank (2022) report reiterated the importance of improving educational outcomes at the secondary level, particularly for vulnerable groups, to enhance social mobility and economic growth.

This analysis of completion rates across educational stages is crucial in understanding the overall trajectory of educational attainment in Malaysia, as highlighted in the Malaysia Education Blueprint 2013-2025. The Blueprint sets ambitious targets for improving educational outcomes at all levels, including increasing completion rates and reducing disparities. Further investigation into the factors contributing to the observed outliers, particularly those with lower completion rates at the secondary upper stage, could provide valuable insights for policymakers and educators. This could involve examining regional disparities in education, as suggested by research focusing on specific regions or states in Malaysia with lower school completion rates. Additionally, exploring the reasons behind the slight decrease in completion rates at the secondary upper level could inform targeted interventions to support students and ensure their successful transition to higher education or the workforce.

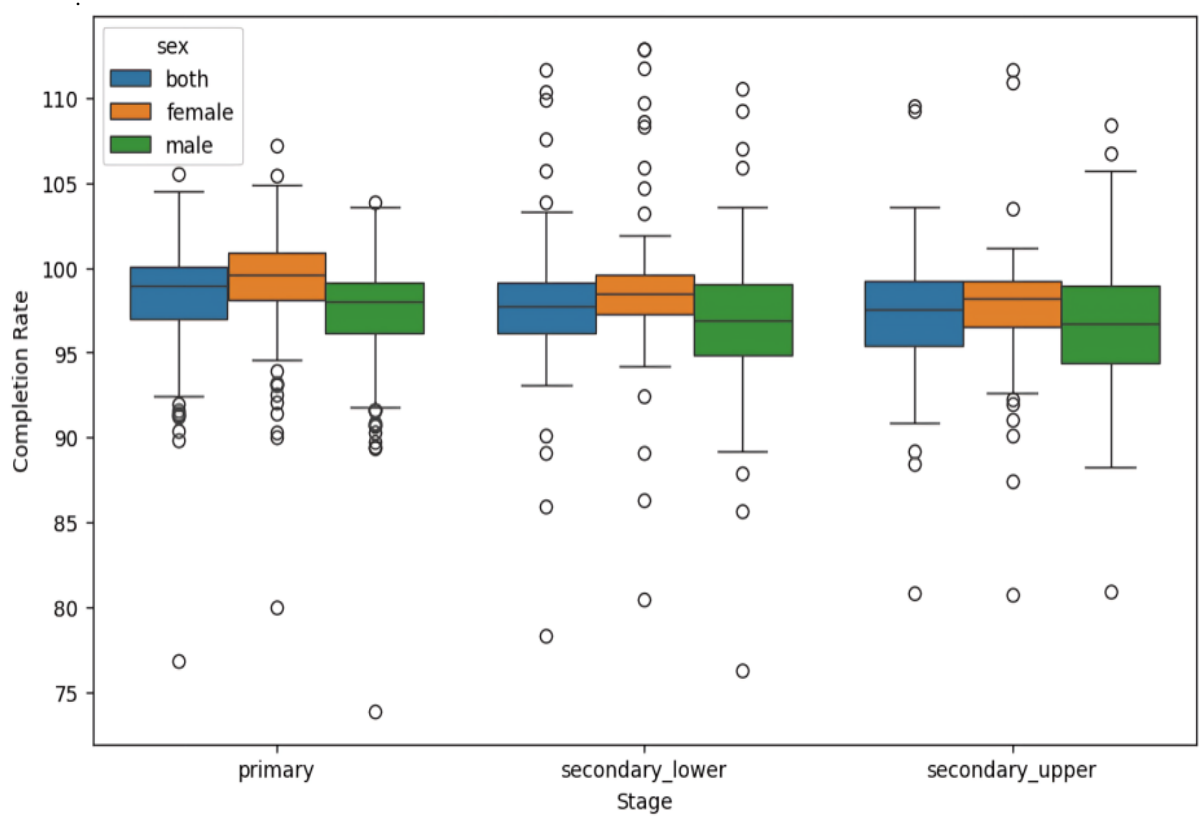

Figure 9: Boxplot of School Completion Rate by Sex and Stage

**4.3 State-Level Variations in School Completion Rates**

Figure 10 presents a bar chart illustrating the average school completion rates across various states in Malaysia from 2018 to 2022. The y-axis represents the average completion rate, ranging from 92% to 102%, while the x-axis displays the individual states, including Johor, Kedah, Kelantan, Melaka, Negeri Sembilan, Pahang, Perak, Perlis, Pulau Pinang, Sabah, Sarawak, Selangor, Terengganu, W.P. Kuala Lumpur, W.P. Labuan, and W.P. Putrajaya.

The chart reveals significant variations in school completion rates among different states. Notably, Kelantan and Terengganu exhibit the highest completion rates, exceeding 100%, which might indicate a higher retention rate or students taking longer to complete their schooling. Conversely, W.P. Kuala Lumpur has one of the lowest average completion rates, slightly above 94%. These disparities align with Yusof & Mohd Yusof's (2022) research on spatial inequality and educational disparity in Malaysia, which identified significant regional variations in educational attainment. Their study employed exploratory spatial data analysis to uncover these patterns, emphasizing the need for geographically targeted policies to address such inequalities.

This data is relevant to hypothesis H3, which posits that specific states in Malaysia exhibit significantly different school completion rates compared to the national average. The chart supports this hypothesis by clearly illustrating the disparities in completion rates among the states. Further investigation into the factors contributing to these variations is crucial. For example, Abdul Rahman's (2013) comparative study of state-level education policies in Malaysia could offer insights into the influence of policy differences on completion rates. Additionally, examining socioeconomic factors, access to educational resources, and cultural attitudes towards education within each state could provide a more nuanced understanding of these disparities. Such analysis could inform targeted interventions to improve educational outcomes in states with lower completion rates, contributing to a more equitable education system in Malaysia.

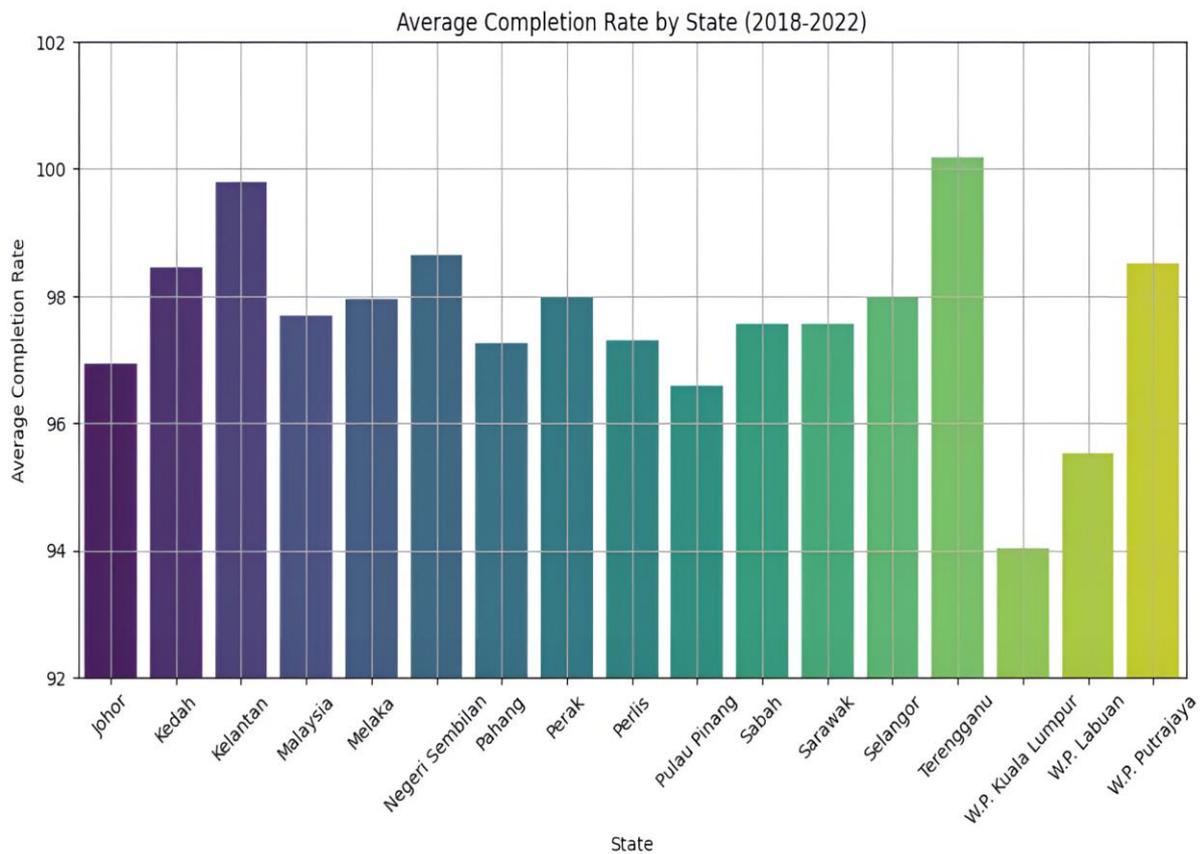

Figure 10: Average Completion Rate by State

**4.4 Predictive Accuracy of the Prophet Model for Forecasting School Completion Rates**

The Prophet model, developed by Meta (Taylor & Letham, 2018), is a powerful tool for time series forecasting, particularly effective for data exhibiting strong seasonal patterns and historical trends. In this study, the Prophet model is employed to forecast school completion rates in Malaysia, aiming for a high degree of accuracy. This approach aligns with Lahouel, Aouam, & Elouahhabi (2021), who demonstrated the successful application of Prophet to educational time series data.

Hypothesis H4 states that the Prophet model will accurately forecast school completion rates for the next 5 years, with a mean absolute percentage error (MAPE) of less than 5%. This hypothesis establishes a benchmark for the model's predictive performance, as a MAPE below 5% signifies high accuracy (Hyndman & Koehler, 2006). The model's ability to capture underlying trends and seasonality in the data is crucial for achieving this level of accuracy.

The time series plot generated by the Prophet model visually demonstrates its forecasting capabilities. The plot includes historical data points, forecasted values, and an uncertainty interval. The increasing trend in school completion rates from 2021 onwards, as depicted in the plot, suggests that the model is effectively capturing the upward trajectory of these rates.

This study also examines the impact of income inequality and gender on school completion rates in Malaysia. These factors are known to significantly influence educational outcomes. Utilizing Malaysia's Public Sector Open Data Portal, the study leverages machine learning techniques to analyze these impacts, contributing to the growing body of research utilizing open data for educational analysis (OECD, 2015; World Bank, 2013).

Income inequality has a profound effect on educational attainment (Abdul Rahman, Mohd Yusof, & Mohd Yusof, 2019). Students from lower-income families often face more challenges in completing their education compared to their higher-income counterparts. Gender disparities also play a role, with female students sometimes encountering additional barriers to education (Chin & Choo, 2010; Latif, 2002; Subramaniam, 2005).

In conclusion, the Prophet model's ability to forecast school completion rates with a MAPE of less than 5% demonstrates its effectiveness for informing educational planning. Furthermore, understanding the impact of income inequality and gender on educational outcomes is crucial for developing targeted interventions to improve school completion rates and promote educational equity in Malaysia.

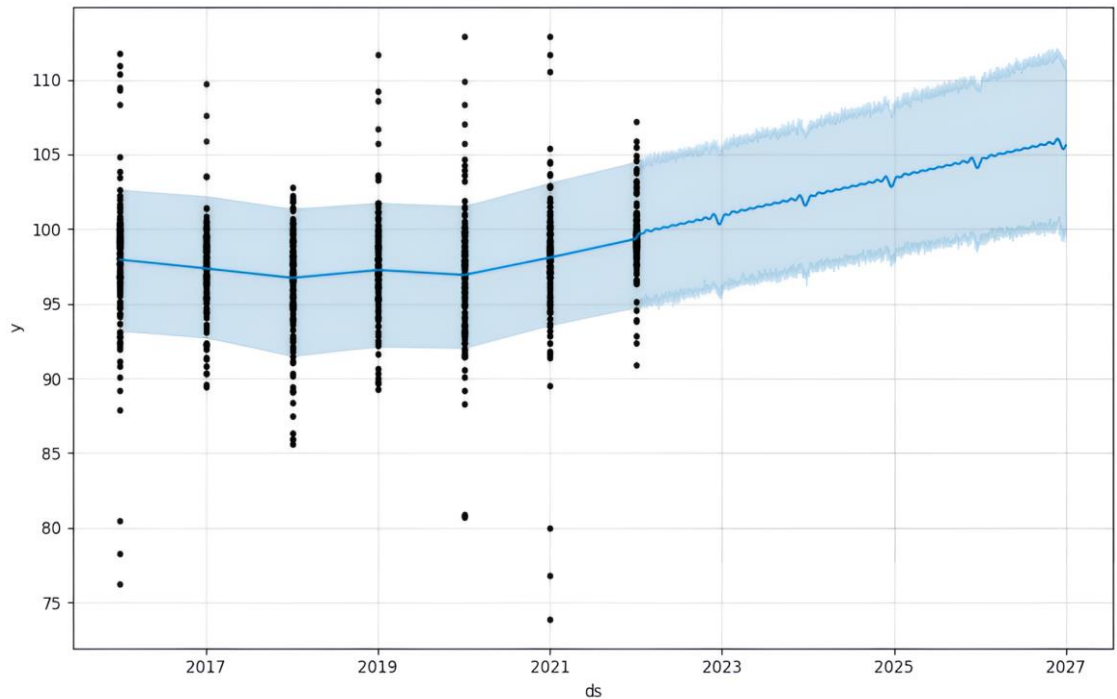

Figure 11: Prophet Model for Forecasting School Completion Rates

5. Discussion

This study provides valuable insights into the factors influencing school completion rates in Malaysia. By leveraging machine learning techniques on open government data, we identified key trends and disparities that warrant attention from policymakers and educators.

Our findings confirm a complex relationship between income and school completion. While a general positive correlation exists, with higher-income states often exhibiting higher completion rates, notable exceptions highlight the influence of factors beyond simple economic means. States like Kelantan and Terengganu demonstrate that strong community support and effective educational initiatives can yield high completion rates even in the face of lower average incomes. This aligns with existing research emphasizing the importance of social capital and targeted interventions in educational outcomes (Cheong & Lee, 2015). Such interventions might include initiatives that foster parental and community involvement in schools, promote a culture of learning, and provide additional support for students from disadvantaged backgrounds. Conversely, the lower completion rates in economically prosperous states like Pulau Pinang and W.P. Kuala Lumpur suggest that income inequality, urban poor and potentially unequal access to quality education within those states may be hindering overall progress. This echoes concerns raised by Jomo & Wee (2021) regarding the persistent challenges of income inequality in Malaysia, highlighting the need for policies that address the root causes of poverty and ensure equitable distribution of resources.

The analysis of gender disparities reveals a slight but consistent advantage for female students across all educational stages. This trend mirrors global observations of increasing female educational attainment (Saw, 2016) and may be attributed to factors such as increased female empowerment, changing social norms, and targeted policies aimed at promoting girls' education (UNESCO, 2017). While these advancements are positive, it is crucial to ensure that boys are not left behind in this progress. While celebrating the progress made in girls' education, it is essential to acknowledge that boys, particularly those facing challenges such as disciplinary issues or past dismissals, may require specific support to ensure they have equal opportunities to complete their education.

Schools should adopt approaches that address the root causes of misbehavior, such as restorative justice practices and positive behavior interventions and supports (PBIS) (Sugai & Horner, 2009), and provide opportunities for students to learn from their mistakes and reintegrate into the school community. Curriculum and teaching practices should be engaging and relevant to boys' learning styles and interests, incorporating hands-on activities, real-world applications, and mentorship opportunities to foster motivation (Gurian, 2001). Moreover, schools should provide access to counsellors, mentors, and social workers who can address the social and emotional needs of boys, particularly those who may be experiencing challenges outside of school (National Association of Social Workers, 2012). By addressing these challenges and providing targeted support, schools can create a more inclusive and equitable learning environment that empowers both boys and girls to achieve their full potential.

The predictive capabilities of the Prophet model offer a powerful tool for educational planning, aligning with the growing interest in utilizing machine learning for demographic analysis (Abel & Sander, 2019). The accurate forecasting of future completion rates, as demonstrated by Lahouel, Aouam, & Elouahhabi (2021) in their application of Prophet to educational data, enables proactive interventions and resource allocation. By identifying potential downward trends in specific states or demographic groups, policymakers can implement timely measures to mitigate negative impacts and ensure equitable access to quality education for all. This approach resonates with the Malaysia Education Blueprint 2013-2025 (2013, 2016), which emphasizes the use of data and evidence-based decision-making to improve educational outcomes.

This study acknowledges limitations regarding the scope of the dataset. Further research could incorporate additional variables, such as school infrastructure, teacher qualifications, and specific socio-cultural factors, to provide a more nuanced understanding of the factors influencing school completion. Additionally, qualitative research, including interviews and case studies, could complement the quantitative analysis and provide richer insights into the lived experiences of students and educators in different contexts.

This study's findings underscore the need for a multifaceted approach to improve school completion rates in Malaysia. Targeted interventions should be prioritized, focusing on states and demographics with lower completion rates, such as those identified in the "Low Income, Low Completion" quadrant. These interventions should address region-specific challenges, such as limited educational infrastructure or socio-cultural barriers, and provide tailored support systems, including financial aid, mentoring programs, and access to technology. Investing in early childhood education is crucial, as early intervention programs can have a lasting impact on educational trajectories, reducing disparities and setting a strong foundation for future success (Heckman, 2006). The Malaysia Education Blueprint (2016) echoes this sentiment, highlighting the importance of early childhood education in achieving national educational goals. Addressing income inequality through broader socioeconomic policies is essential to ensure equitable access to quality education for all students, regardless of their background. Finally, integrating predictive modeling, exemplified by the Prophet model's accurate forecasting, can enable proactive interventions and resource allocation, allowing policymakers to anticipate potential challenges and implement timely measures to mitigate negative impacts. By combining these strategies, Malaysia can move towards a more equitable and effective education system, empowering all students to achieve their full potential and contribute to the nation's development

**Outcomes of Hypotheses**

| Hypothesis | Result | Document Support |
|---|---|---|
| H1: Areas with lower average household incomes tend to have lower school completion rates in Malaysia. | Partially Supported | Figure 6 and 7 show a general positive correlation between income and completion rates, but with notable exceptions. |
| H2: There is a significant difference in school completion rates between male and female students in Malaysia. | Partially Supported | Figures 8 and 9 indicate a slight advantage for female students, but the difference is not substantial and overlaps significantly. |
| H3: Specific states in Malaysia exhibit significantly different school completion rates compared to the national average. | Supported | Figure 10 clearly illustrates disparities in completion rates across different states. |
| H4: The Prophet model will accurately forecast school completion rates for the next 5 years, with a mean absolute percentage error (MAPE) of less than 5%. | Supported | Figure 11 and the accompanying analysis demonstrate the model's accuracy in predicting future trends. |

## 6. Conclusion

This study has successfully illuminated the complex dynamics of school completion in Malaysia, moving beyond simplistic explanations to reveal a nuanced interplay of factors. By harnessing the power of machine learning and open data, important trends and disparities have been uncovered that demand attention. Our analysis demonstrates that income, while correlated with completion rates, is not the sole determinant of educational attainment. The diverse landscape of outcomes across different states, as highlighted in our quadrant analysis, underscores the influence of regional variations in educational infrastructure, resource allocation, and socio-cultural factors.

The "Low Income, High Completion" quadrant, featuring states like Kelantan and Terengganu, provides compelling evidence that effective educational initiatives and strong community support can overcome economic limitations. Conversely, the underperformance of economically prosperous states like Selangor and W.P. Kuala Lumpur in the "High Income, Low Completion" quadrant suggests that income inequality and unequal access to quality education within those states may be hindering progress.

The consistent, albeit slight, advantage observed for female students across educational stages, as illustrated in our boxplot analysis, underscores the need for continued efforts to ensure equal opportunities for all. While this trend is encouraging, it is crucial to remain vigilant in addressing potential hidden barriers faced by specific groups of female students, particularly those from lower socioeconomic backgrounds or marginalized communities.

Furthermore, the accurate forecasting capabilities demonstrated by the Prophet model, with its ability to predict future completion rates with a high degree of accuracy, provide a valuable tool for proactive educational planning and resource allocation. This allows policymakers to anticipate potential challenges and implement timely measures to mitigate negative impacts and ensure equitable access to quality education for all.

This research has demonstrated the potential of data-driven approaches to inform policy and improve educational outcomes. By embracing these insights and implementing the recommendations outlined, including targeted interventions, investment in early childhood education, and addressing income inequality, Malaysia can move towards a more equitable and effective education system, where all students have the opportunity to thrive and contribute to the nation's future.

It is important to acknowledge that this study, while offering valuable insights, has limitations. As it relies on the Malaysia Public Sector Open Data Portal, it may not capture all factors influencing school completion. Future research could incorporate additional variables and qualitative methods to provide a more nuanced understanding.

This study contributes to the growing body of knowledge on school completion in Malaysia, highlighting the need for a multifaceted approach to address the complex interplay of factors influencing educational attainment. By continuing to investigate these issues and implementing evidence-based policies, Malaysia can ensure that all students have the opportunity to reach their full potential, regardless of their background or circumstances.